\documentstyle[12pt]{article}
\begin{document}
\begin{titlepage}
\rightline{SNUTP/97-040}
\def\today{\ifcase\month\or
        January\or February\or March\or April\or May\or June\or
        July\or August\or September\or October\or November\or December\fi,
  \number\year}
%\rightline{\today}
\rightline{hep-th/9703171}
\rightline{Feb., 1997}
\vskip 1cm
\centerline{\LARGE Instanton-induced Effective Vertex in the}
\vskip 0.5cm
\centerline{\LARGE Seiberg-Witten Theory with Matter}
\vskip 2cm
\centerline{\sc ByungKoo Lee \footnote{bklee@nms.kyunghee.ac.kr}
 and Soonkeon Nam \footnote{nam@nms.kyunghee.ac.kr}}
\vskip 1cm
\centerline {{\it Department of Physics and}}
\centerline {{\it Research Institute for Basic Sciences,}}
\centerline {{\it Kyung Hee University,}}
\centerline {{\it Seoul 130-701, Korea}}
\vskip 1cm
\centerline{\sc Abstract}
\vskip 0.2in
The instanton-induced effective vertex is derived for $N=2$
supersymmetric QCD (SQCD) with arbitrary mass matter hypermultiplets
for the case of $SU(2)$.
The leading term of the low energy effective lagrangian
obtained from this vertex agrees with
one-instanton effective term of the Seiberg-Witten result.

%PACS number(s):
\end{titlepage}
\newpage

\def\beq{\begin{equation}}
\def\eeq{\end{equation}}
\def\bea{\begin{eqnarray}}
\def\eea{\end{eqnarray}}
\renewcommand{\arraystretch}{1.5}
\def\ba{\begin{array}}
\def\ea{\end{array}}
\def\bce{\begin{center}}
\def\ece{\end{center}}
\def\nn{\noindent}
\def\nonu{\nonumber}
\def\pbx{\partial_x}
\def\prt{\partial}
\def\n{\newline\indent}
%%%%%%%%%%%%%%%%%%%

\newcommand{\sibsi}{\sigma_{\mu}\bar{\sigma_{\nu}}}
\newcommand{\bsisi}{\bar{\sigma_{\mu}}\sigma_{\nu}}
\newcommand{\F}{F_{\mu\nu}}
\newcommand{\al}{\alpha}
\newcommand{\dal}{\dot{\alpha}}
\newcommand{\bt}{\beta}
\newcommand{\ep}{\epsilon}
\newcommand{\dl}{\delta}
\newcommand{\bep}{\bar{\epsilon}}
\newcommand{\bsi}{\bar{\sigma}}
\newcommand{\bbt}{\bar{\beta}}
\newcommand{\bla}{\bar{\lambda}}
\newcommand{\bth}{\bar{\theta}}
\newcommand{\bet}{\bar{\eta}}
\newcommand{\bPh}{\bar{\Phi}}
\newcommand{\bph}{\bar{\phi}}
\newcommand{\bQ}{\bar{Q}}
\newcommand{\bPsi}{\bar{\Psi}}
\newcommand{\si}{\sigma}
\newcommand{\nbl}{\nabla}
\newcommand{\qnbl}{\nabla^{2}}
\newcommand{\bnbl}{\bar{\nabla}}
\newcommand{\qbnb}{\bar{\nabla}^{2}}
\newcommand{\tht}{\theta}
\newcommand{\qpi}{\pi^{2}}
\newcommand{\qrho}{\rho^{2}}
\newcommand{\qd}{d^{2}}
\newcommand{\td}{d^{3}}
\newcommand{\fd}{d^{4}}
\newcommand{\qg}{g^{2}}
\newcommand{\la}{\lambda}
\newcommand{\La}{\Lambda}
\newcommand{\dbt}{\dot{\beta}}
\newcommand{\nd}[1]{/\hspace{-0.5em} #1}
%%%%%%%%%%%%%%%%%%%%%%%%%%

\def\lbr{\left(}
\def\rbr{\right)}
\def\half{\frac{1}{2}}

\def\vol#1{{\bf #1}}
\def\nupha#1{Nucl. Phys. \vol{#1} }
\def\CMP#1{Comm. Math. Phys. \vol{#1} }
\def\phlta#1{Phys. Lett. \vol{#1} }
\def\phyrv#1{Phys. Rev. \vol{#1} }
\def\PRL#1{Phys. Rev. Lett \vol{#1} }
\def\prs#1{Proc. Roc. Soc. \vol{#1} }
\def\PTP#1{Prog. Theo. Phys. \vol{#1} }
\def\SJNP#1{Sov. J. Nucl. Phys. \vol{#1} }
\def\TMP#1{Theor. Math. Phys. \vol{#1} }
\def\ANNPHY#1{Annals of Phys. \vol{#1} }
\def\PNAS#1{Proc. Natl. Acad. Sci. USA \vol{#1} }

During the past year there has been a lot of progress in $N=2$ supersymmetric
Yang-Mills theories in four dimensions. Using the idea of duality and holomorphy, Seiberg and
Witten determined the exact low energy effective lagrangian for the gauge
group $SU(2)$ without \cite{SW} and with matter hypermultiplets \cite{SW2}.
This low energy effective lagrangian is determined by a single holomorphic
function: the prepotential $\cal F$.
By carefully studying of its singular structures in moduli space, 
they determined not only the form  but the numerical cefficients
of the prepotential exactly.\n
The moduli spaces of these theories are described by 
hyper elliptic curves and can be related to integrable models \cite{G}.
One of the most interesting feature of this prepotential is that it contains an
infinite series of instanton contributions \cite{S}.
Noticing this important feature several authors tried to check the Seiberg
-Witten result with semi-classical instanton calculation at weak coupling limit
without using any duality conjecture \cite{FP}.
This direct microscopic instanton calculation for $N=2$ $SU(2)$ SUSY Yang-Mills
theory provides a nontrivial check of the idea of duality and has been
carried out at the one-instanton \cite{FP}, and two-instanton 
levels using the ADHM construction \cite{F,DKM}.\n
Shortly thereafter multi-instanton calculation for SUSY Yang-Mills 
theory coupling to matter has been performed by the two independent
groups \cite{DKM,AHSW}.
They found that there is an excellent agreement between the Seiberg-Witten 
result and the semi-classical instanton calculation except for 
the $N_f=3,4$ cases.
At the one instanton level this microscopic calculation has been extented to
the group $SU(N)$, again with and without matter \cite{IS,IS2} and also to the
semi-simple Lie groups \cite{IS3}.\n
Another direction of studying the instanton effects in the low energy 
effective lagrangian has been suggested by Yung \cite{Y}.
In that approach the nonperturbative instanton effect was represented,
according to the perturbation theory language, as a four fermion vertex
attached to the tree level lagrangian  and one can derive one instanton-induced
effective vertex and find that in the low energy limit the leading term
coincides with the Seiberg-Witten effective action.
This provides get another nontrivial check on the exact results. 
\n
In this letter we consider the one instanton-induced  effective vertex for
$N=2$ $SU(2)$ SUSY gauge theory with matter hypermultiplets which have 
arbitrary masses. 
Experience from the Seiberg-Witten theory tells us that simple 
addition of matter can lead to quite different structure compare 
to the pure Yang-Mills case \cite{SW2}. \n
The model we are considering is the $N=2$ SQCD which has an $N=1$ chiral
multiplet ${\Phi}=(\phi,\psi)$ in the adjoint representation of the group
$SU(2)$ and $N=1$ vector multiplet $W_{\alpha}=(\la,v_{\mu})$, which form
an $N=2$ vector multiplet. There are also $N=1$ chiral multiplets $Q_{k}=
(q_{k},\psi_{mk})$ and $\tilde{Q}_{k}=(\tilde{q}_{k},\tilde{\psi}_{mk})\,
(k=1,\cdots,N_{f})$, which form the $N=2$ matter hypermultiplets
in the fundamental representation of the group.
There exist global $SU(2)_{R}$ under which
the superfields transform as follows:
\beq
\la \leftrightarrow \psi,\,\,q\rightarrow \tilde {q}^{\dagger},\,\,
\tilde{q}\rightarrow\ -q^{\dagger}, 
\eeq
while gauge and scalar fields are singlets under the transformation. \n
The lagrangian of the model is given by
\beq
{\cal L}_{SQCD} = {\cal L}_{SYM} +{\cal L}_{matter} +{\cal L}_{Yukawa}
+{\cal L}_{mass},
\eeq
where each term is given as follows:
\bea
{\cal L}_{SYM} & =  &
\frac{1}{4g^2}\int d^2 \theta W_{\alpha}^a W^{\alpha a}
+\frac{1}{4g^2}\int d^2 \bar{\theta}
\bar{ W}^{\dot{\alpha}a}\bar{W}_{\dot{\alpha}}^{a},\\
{\cal L}_{matter} & =  &
\int d^2 \theta d^2\bar{\theta} \left[{\Phi}^{\dagger a}
(e^{-2gV}\Phi)^a
+\sum_{k=1}^{N_f}(Q_{k}^{\dagger}e^{-2gV} Q_{k}
+\tilde{Q}_{k}e^{2gV}\tilde{Q}^{\dagger}_{k}) \label{M}
\right], \\
{\cal L}_{Yukawa} & =  &
i{\sqrt2} g\int d^2 \theta \sum_{k=1}^{N_f}\tilde{Q}_{k}\Phi 
Q_{k},\\
{\cal L}_{mass} & =  & 
\int d^2 \theta \sum_{k=1}^{N_f} m_{k}\tilde{Q}_{k}Q_{k}.
\eea
In the above $\al$ is the Weyl index, $a$ is the color index in the
adjoint representation of the group $SU(2)$, and $g$ is the gauge coupling
constant. 
The first term on the RHS of eq.(\ref{M}) is the adjoint representation
and the second term is the fundamental representation of the matter.\n
The supersymmetric generalization of the instanton has been studied in
the context of dynamical SUSY breaking by instanton effect
with the hope that it solves the gauge hierarchy problem \cite{ADS,AK}.
The defining equation of SQCD instanton in the weak coupling 
limit, up to the leading order of $g$, consists of
the vector multiplets with the following 
equation of motion \cite{DKM,DKM2,AHSW};
\bea
F_{\mu\nu}=-\tilde{F}_{\mu\nu}, \hspace{2.2cm} \\
\nd{\bar{D}}\la=0, \quad \nd{\bar{D}}\psi=0, \quad
D^2\phi-i\sqrt{2}[\lambda,\psi]=0,
\eea
and the hypermultiplets satisfying the following equation of motion;
\bea
\nd{\bar{D}}\psi_{m}=0, \qquad \nd{\bar{D}}\tilde{\psi}_m = 0,
\hspace{5em} \\
D^2 q-i\sqrt{2}g\la\psi_{m}=0, \,\,\qquad
D^2\tilde{q}+i\sqrt{2}g\tilde{\psi}_{m}\la=0, \,\hspace{0.8cm}\\
D^2 q^{\dagger}-i\sqrt{2}g\tilde{\psi}_{m}\psi=0, \qquad
D^2\tilde{q}^{\dagger}-i\sqrt{2}g\psi{\psi}_{m}=0. \hspace{0.8cm}
\eea
In supersymmetric theory, due to the cancellation between bosonic 
and fermionic excitation modes around classical instanton background, the
instanton measure takes rather a simple form that depends only on the
zero mode contributions.
For bosonic case there are eight zero-modes that correspond to 
the translation, isorotation and scale transformation.
The measure takes the form \cite{F,Y,T,ADS};
\beq
\big[d\mu_{boson}\big]=
2^{10}\pi^{6}M^{8}\rho^{8}\exp\left(-\frac{8\pi^2}{g^2}\right)
d^{4}x_{0}\left(\frac{d\rho}{\rho^{5}}\right)\left(\frac{\td u}{2\qpi}
\right),
\eeq
where $\rho$ dependence has been properly chosen to satisfy the integration
measure does not have any mass dimension.
There is a contribution to the measure that comes from adjoint fermion
zero modes. 
Under the proper normalization it has the form \cite{Y}
\beq      
\big[d\mu_{fermion}\big]=
  \frac{1}{16\qpi M}\qd\al\,\,\frac{1}{16\qpi M}\qd\zeta\,\,
  \frac{1}{32\qpi\rho^{2} M}\qd\bbt\,\,\frac{1}{4\qpi\qrho v^{2} 
  M}\qd\bet,
\eeq
where $\al, \zeta$ denote the supersymmetric mode of gaugino
and higgsino, while $\bbt, \bet$ denote the superconformal mode of
the gaugino and higgsino, respectively.
In addition there are another contribution comes from
hypermultiplet matter zero modes \cite{DKM2,AHSW}.
\beq
\big[d\mu_{hyp}\big]=
\pi^{-2N_f} d^{N_{f}}\xi d^{N_{f}}\tilde{\xi} \equiv
\pi^{-2N_f}\prod_{k=1}^{N_f} d^{N_{f}}\xi_{k} 
d^{N_{f}}\tilde{\xi}_{k},
\eeq
where the Grassmann variables $\xi_k$ and $\tilde\xi_k$ denotes
the $k$-th hypermultiplet
zero mode in the fundamental representation of the gauge group $SU(2)$.\n
To obtain full SQCD instanton measure,
we should combine these three and we have
\bea \label{N2}
\big[d\mu_{SQCD}\big] & = & 
\big[d\mu_{boson}\big]\big[d\mu_{fermion}\big]\big[d\mu_{hyp}\big]\\ 
              & = & \frac{1}{32\pi^{2+2N_f}}\frac{\La^{4-N_f}}{v^{2}}\fd x_{0}
\frac{d\rho}{\rho}\frac{d^{3}
u_{inv}}{2\qpi}\qd\al\qd\zeta\qd\bbt_{inv}\qd\bet_{1}
d^{N_f}\xi_{k} d^{N_{f}}\tilde{\xi}_{k},
\eea
where the invariant orientation is defined by \cite{Y}
\beq
  u^{\al\dal}_{inv}= u^{\al\dbt}\exp\big[-4i\bbt_{\dbt}\bet^{\dal} -
 2i\dl^{\dal}_{\dbt}(\bet\bbt)\big],
\eeq
and SUSY invariant collective coordinates are
\beq
\bbt_{inv}=\bbt(1+4i\bbt\bet),
\eeq
\beq
\bar{\eta}_{1}= \frac{\bar{\eta}}{1+4i\bar{\bt}\bar{\eta}},
\eeq
and $\La = M \exp(-\frac{8\pi^2}{g^2})$ is a dynamical scale
that depends on the regularization scheme.
Here we use the Pauli-Villars regularization scheme, which is
normally accepted in the instanton calculation.
It is well known that in supersymmetric theory the
$\beta$-function is only one loop effect and does not receive any 
higher order contribution \cite{S} and
the power change of $\La$ can be expected from the behaviour of one-loop
$\beta$-function coeffient when the matter fields are present.
To satisfy the asymptotic freedom, we
restrict the value of $N_f$ as the integer runs from 0 to 4.
It can be readily checked that the measure has correct mass 
dimension zero. \n
By the way it is known that the instanton measure for $N=1$
SUSY case can be fixed only by supersymmetry and dimensional
arguments \cite{NSVZ} and it takes the form
\beq
\big[d\mu_{N=1}\big]\exp(-S) = C\frac{\La^{4}}{v^{2}}\exp\left(-\frac{4\pi^2}
{g^2}\rho^2_{inv}v^2\right)\fd x_{0}\frac{d\rho}{\rho}\frac{d^{3}
u_{inv}}{2\qpi}\qd\tht_{0}\qd\bth_{0}\qd\bbt\qd\bet,
\label{N1}
\eeq
where $C$ is the numerical constant that depends on the definition of
the scale parameter $\La$ and the Grassmann variables $\tht_{0}, \bth_{0}$
denote the supersymmetric collective coordinates and the matter fermion
zero mode, repectively.
Comparing eq.(\ref{N1}) with eq.(\ref{N2})
we see that $N=1$ instanton measure has very similar form as that 
of $N=2$. This observation leads us to use the $N=1$ results for
$N=2$ SUSY case.\n
In $N=2$ SUSY, as already mentioned, the hypermultiplet
is described in the $N=1$ chiral multiplet $Q$ and an anti-chiral
multiplet $\tilde{Q}^\dagger$, both transform as $N$ representation of
the gauge group $SU(N)$. There are another multiplets $\tilde{Q}$ and
$Q^\dagger$ which transform as $\bar{N}$ representation of the group. \n
When the scalar matter fields have vanishing vacuum expectation values
(VEV's), the matter multiplets in the one-instanton background
have the following form \cite{N,ADS,AK}
\beq
Q^{i}_{k} = \tht^{\al}(\psi_{m \al})^{i}_{k}=
-\tht^{i}\xi_{k}\frac{1}{\rho^2 f^{3/2}}, \hspace{1cm}
{Q}^{\dagger}_{ik} =\tht^{\al}({\psi}_{m \al}^{\dagger})_{ik}=
-\tht_{i}{\bar\xi}_{k}\frac{1}{\rho^2 f^{3/2}},
\eeq
and
\beq
{Q}^\dagger_{k}{Q}_{k} = -\tht^{2}\xi_{k}{\bar\xi}_{k}\frac{1}{\rho^4 f^3},
\label{QQ} \label{V}
\eeq
where
\beq
f = 1 + \frac{x^2}{\rho^2},
\eeq
where $i$ denotes an isospin and $k$ denotes the flavor. \n
For the case of the squarks have nonvanishing VEV's
the product becomes \cite{NSVZ,N}
\beq
{Q}^\dagger_{k}Q_{k} = \frac{\tilde x^2 q^\dagger_{k}q_{k}}{\tilde x^2 +\rho^2}
\label{Q},
\eeq
where
\beq
\tilde x_{\al \dal} = (x-x_{0})_{\al \dal}+ 
4i\tilde{\tht}_{\al}(\bth_{0})_{\dal},
\eeq
\beq
\tilde{\tht}_{\al}=(\tht-\tht_{0})_{\al}+(x-x_{0})_{\al\dal}\bbt^{\dal},
\eeq
and $q_{k}, q^\dagger_{k}$ are the VEV's of the $k$-th hypermultiplets.
In fact in this case the instanton solution does not exist, but
instanton-like field configuration can be obtained by the method
of constrained instanton \cite{ADS,A}. \n
In the large distance limit i.e. $x\gg\rho$, eq.(\ref{V}) and 
eq.(\ref{Q}) takes the form
\beq
{Q}^\dagger_{k}Q_{k}=\left\{ \begin{array}{ll}
\displaystyle{ -\tht^{2}\xi_{k}\bar{\xi}_{k}\frac{\rho^2}{\tilde x^6}}
& \hspace{3em}\mbox{for $q_{k},\,q^\dagger_{k}=0$},  \\
q^\dagger_{k}q_{k}\left(1-\displaystyle{\frac{\rho^2}{\tilde x^2}}
\right)
& \hspace{3em}\mbox{for $q_{k},\, q^\dagger_{k}\neq 0$}. \end{array} \right.
\eeq
From the eq.'s (\ref{QQ}), (\ref{Q}),
the calculation of $S_{mass}$ becomes \cite{FS}
\beq 
S_{mass}=
\left\{ \begin{array}{ll}
-\pi^{2}\displaystyle{\sum_{k=1}^{N_f}m_{k}} \tilde{\xi}_{k}\xi_{k} &
\hspace{3em}\mbox{for $q_{k},\,{\tilde q_k}=0$},  \\
-16\pi^2\rho^2\displaystyle{\sum_{k=1}^{N_f}m_{k}} q_{k}\tilde q_{k}
 \tilde{\xi}_{k}
 \xi_{k} & \hspace{3em}\mbox{for $q_{k},\,{\tilde q_k}\neq 0$},\end{array}
\right. 
\eeq
and it gives the additional contribution to the instanton measure. \n
This pattern already has been shown for $N=1$ SUSY case \cite{NSVZ},
and also
for $N=2$ case in the Coulomb branch where the squarks have 
vanishing VEV's \cite{DKM2,AHSW}.
These mass contributions lead to the change of prepotential that
agrees with the previous result obtained by Ohta solving 
the Picard-Fuchs equations for $N=2$ $SU(2)$ massive Yang-Mills
theory \cite{O}. \n
Now let us consider the instanton-induced effective vertex, which 
is our main concern.
Inspired by the work in Ref.\cite{CDG}, an
instanton-induced effective vertex for $N=2$ SUSY Yang-Mills theory
has been derived, and
has been shown that in the large distance limit the correlation  function
contracted with this vertex yields the original $N=2$
instanton superfield \cite{Y}. \n
Considering the correlation function of the hypermultiplet matter fields
in the instanton background,
\beq
\langle Q_{i}(x^{L}_{1},\tht_{1})\,{Q}^{\dagger}_{j}(x^{R}_{2},\bth_{2})
\,{\tilde {Q}}^{\dagger}_{k}(x^{R}_{3},\bth_{3})\,\tilde{Q}_{l}
(x^{L}_{4},\tht_{4})\rangle,
\eeq
it can also be expressed as an average over quantum fluctuations around
perturbative vacuum by inserting the instanton induced effective vertex.
The vertex to be added to the classical action takes the form
$\exp(-V_{I})=\sum_{k}\frac{1}{k!}(-V_{I})^{k}$, where $k$-th term corresponds
to the $k$-instanton contribution.
Inserting the vertex and let us consider only the one-instanton
contribution: $(-V_{I}^{N=2})$. 
When the squarks have vanishing VEV's, the product becomes
\beq
{Q}^\dagger_{k}{Q}_{k} = -\tht^{2}\xi_{k}\bar{\xi}_{k}\psi^\dagger_{mk}
\psi_{mk}+\cdots,
\eeq
and when the squarks have nonvanishing VEV's,
expanding $Q^\dagger$ and $Q$ around their expectation values, we have
\beq      
 Q^{\dagger}_{k}Q_{k} = q^\dagger_{k}q_{k} + q^\dagger_k\delta
 Q+\delta {Q}^{\dagger}q_{k}+
 O(\delta Q^{\dagger}\delta Q),
\eeq
where we dropped out the term quadratic in quantum fluctuations.
Using the propagator for hypermultiplet matter fields,
\beq      
\langle Q_{i}(x^{L}_{1},\tht_{1})\,\delta Q^{\dagger}_{j}(x^{R}_{0},\bth)
\rangle =
\frac{\delta_{ij}g^{2}}{(2\pi)^{2}}e^{-2i\bth\nd{\bar{\prt}}\tht_{1}}
\frac{1}{(x^{L}_{1}- x^{R}_{0})^{2}},
\eeq
and the propagator for the Weyl fermion, 
\beq
\langle\psi_{mi}(x^{L}_{1},\tht_{1})\,{\bar\psi}_{mj}(x^{R}_{0},\bth)
\rangle =
\frac{\delta_{ij}g^{2}}{(2\pi)^{2}}
e^{-2i\bth\nd{\bar{\prt}}\tht_{1}}{\nd\prt}
\frac{1}{(x_1^L-x_0^R)^{2}},
\eeq
where the left and the right coordinates are defined as
\beq
x^{L}_{\mu}=x_{\mu}+i(\bth\bar{\sigma}_{\mu}\tht),\hspace{1cm}
x^{R}_{\mu}=x_{\mu}-i(\bth\bar{\sigma}_{\mu}\tht),
\eeq
we can deduce the form of instanton induced effective vertex. \n
When the squarks have the vanishing VEV's,
the resulting $N=2$ SQCD vertex in terms of $N=1$ supermultiplets
is as follows:
\bea
\lefteqn{V^{N=2}_{I} = -\frac{1}{32\pi^{2+2N_f}}\La^{4-N_f}\int d^{4}x_{0}
 \frac{d\rho}{\rho}\frac{d^{3}u_{inv}}{2\pi^{2}}\frac{1}{\Phi^{a2}}
 d^{2}\al d^{2}\zeta d^{2}\bbt_{inv}d^{2}\bet_{1}d^{N_{f}}\xi d^{N_{f}}
 \tilde{\xi}} \nonu \\
 & & \exp \left[-\frac{4\pi^{2}}{g^{2}}\rho^{2}_{inv}\bPh^{a}\Phi^{a}
 +\left(\frac{4\pi^{2}}{g^{2}}\right)^2\rho^{2}_{inv}
 \sum_{k=1}^{N_f}({Q}^{\dagger}_{k}Q_{k}+\tilde{Q}^\dagger_{k}\tilde{Q}_{k})
 -\frac{4\sqrt{2}\pi^{2}}{g^{2}}\rho^{2}_{inv}\bPh^{a}W^{a}_{\al}
 \zeta^{\al} \right. \nonumber  \\
 & &\left. +\,\,\frac{2\pi^{2}}{g^{2}}i\rho^{2}_{inv}\bar{\nabla}_{\dbt}
 (\bar{W}^{a}\bar{u}_{inv}\tau^{a}u_{inv})^{\dbt} - \frac{16\pi^{2}}{g^{2}}
 \rho^{2}_{inv}(\bar{W}^{a}\bar{u}_{inv}\tau^{a}u_{inv}\bbt_{inv})
 \right.\nonumber \\
 & &\left. +\,\,\frac{2\sqrt{2}\pi^{2}}{g^{2}}
 \rho^{2}_{inv}(\zeta\nd{D}\bar{u}_{inv}\tau^{a}u_{inv}\bar{\nabla}\bPh^{a})
 -\frac{16\sqrt{2}\pi^{2}}{g^{2}}i
 \rho^{2}_{inv}(\zeta\nd{D}\bar{u}_{inv}\tau^{a}u_{inv}\bbt_{inv})\bPh^{a}
 \nonumber \right. \\
 & & \left.+\,\,\int d^{4}x d^{2}\tht \sum_{k=1}^{N_f} m_{k}\tilde{Q}_{k}Q_{k}
 \right]. 
\eea
Defining $N=2$ chiral superfields as
\beq      
\Psi^{a}(x^{L}_{N=2},\tht_{1},\tht_{2})=\Phi^{a}(x^{L}_{N=2},\tht_{1})+
\sqrt{2}\tht_{2\al}W^{a\al}(x^{L}_{N=2},\tht_{1})+\cdots ,  
\eeq
\beq      
\bPsi^{a}(x^{R}_{N=2},\bth_{1},\bth_{2})=\bPh^{a}(x^{R}_{N=2},\bth_{1})+
\sqrt{2}\bth^{\dal}_{2}\bar{W}^{a}_{\dal}(x^{R}_{N=2},\bth_{1})+\cdots ,  
\eeq 
where $N=2$ coordinates are defined as
\beq
(x^{L}_{\mu})_{N=2}=x_{\mu}+i\bth_{1}\bsi_{\mu}\tht_{1}+
i\bth_{2}\bsi_{\mu}\tht_{2},
\eeq
\beq
(x^{R}_{\mu})_{N=2}=x_{\mu}-i\bth_{1}\bsi_{\mu}\tht_{1}-
i\bth_{2}\bsi_{\mu}\tht_{2}.
\eeq
Expanding field $\Psi_{a}$ around its VEV and considering the 
gauge symmetry breaking
\beq
\Psi_{a}=v\delta_{a3}+\delta\Psi_{a}
\eeq
the $\Psi_{1}$ and $\Psi_{2}$ components become massive and do 
not propagate at large distances \cite{Y}.
Therefore the low energy theory, integrating out these heavy 
fields, depends only massless field $\Psi_{3}$.
Then the resulting low energy effective vertex, in terms of $N=2$
SUSY, has the form
\bea
V^{LE}_{I}& = & -\frac{1}{4\pi^{2+2N_f}}\La^{4-N_f}\int d^4 x_{0}
\frac{d\rho}{\rho}\frac{d^{3} u_{inv}}{2\pi^2}
\frac{1}{\Psi^{4}_{3}}d^{2}\tht_{1} d^2\bth_{1}
d^2\tht_{2}d^{2}\bth_{2}d^{N_{f}}\xi d^{N_{f}}\tilde{\xi}
\nonu \\
& & \exp\left[-\frac{4\pi^2}{g^2}\rho^{2}_{inv}\bar{\Psi}
_{3}\Psi_{3}+\left(\frac{4\pi^2}{g^2}\right)^2\rho_{inv}^{2}
\sum_{k=1}^{N_f}({Q}^\dagger_{k} Q_{k}+\tilde{Q}^{\dagger}_{k}\tilde{Q}_{k})
\right.\nonu \\
& & \left.-\frac{\pi^{2}}{\sqrt{2}g^2}\rho^{2}_{inv}
i(\bar{\nabla}_{i}\bar{u}_{inv}\tau^{3}u_{inv}
\bar{\nabla^{i}})\bar{\Psi}_{3}
+\int d^{4}x d^{2}\tht\sum_{k=1}^{N_f} m_{k}\tilde{Q}_{k}Q_{k}
\right],
\eea
where new Grassmann parameters are introduced
\beq
\tht_{1}=-\al, \,\; \tht_{2}=\zeta, \,\; \bth_{1}=-\bet_{1},
\,\; \bth_{2}=\bet_{2},
\eeq
\beq
\bbt_{inv\dal}=\frac{v}{2\sqrt2}(\bar{u}_{inv}\tau_{3}u_{inv}
\bet_{2})_{\dal},
\eeq
and also we have used the $N=2$ SUSY promotion rule $v\rightarrow
\Psi_3$. \n
For the case of nonvanishing VEV's, the resulting low energy effective
vertex has the form as follows:
\bea
\lefteqn{V^{LE}_{I}=-\frac{\La^{4-N_f}}{64\pi^{2+2N_f}}\int d^4 x_{0}
\frac{d\rho}{\rho^{1+2N_f}}\frac{d^{3} u_{inv}}{2\pi^2}
\frac{1}{\Psi^{4}_{3}}\prod_{k=1}^{N_f} Q_k^{-1}\tilde{Q}_{k}^{-1}d^{2}
\tht_{1} d^2\bth_{1}
d^2\tht_{2}d^{2}\bth_{2}d^{N_{f}}\xi d^{N_{f}}\tilde{\xi}} \nonu \\
& & \exp\left[-\frac{4\pi^2}{g^2}\rho^{2}_{inv}\left(\bar{\Psi}_{3}
\Psi_{3}+\frac{1}{2}
\sum_{k=1}^{N_f}({Q}^\dagger_{k} Q_{k}+\tilde{Q}^\dagger_{k}
\tilde{Q}_{k}) \right)
-\frac{\pi^{2}}{\sqrt{2}g^2}\rho^{2}_{inv}
i(\bar{\nabla}_{i}\bar{u}_{inv}\tau^{3}u_{inv}
\bar{\nabla^{i}})\bar{\Psi}_{3} \right.\nonu \\
& & \left.+\,\, \int d^4 x d^2 \tht \sum_{k=1}^{N_f} m_{k}
\tilde{Q}_{k}Q_{k}\right].
\eea
where we have used the $N=1$ SUSY promotion rule $q_{k}\rightarrow Q_{k}$. 
Using the identity for any function $f$
\beq
 \int \qd\bth_{1}\qd\bth_{2}f(\qrho_{inv})= -\frac{1}{2}\Psi^{2}_{3}
 \rho^{4}\left(\frac{\prt}{\prt\qrho}\right)^{2}f(\qrho),
\eeq
as for the case of without matter hypermultiplet, the integral over
$\rho$ reduced to a total derivative and there is only zero size
instanton contribution \cite{Y,NSVZ}. 
Carrying out the integration we have
\beq
V^{LE}_{I}=\frac{\La^{4-N_f}}{16\pi^2}\prod_{i=1}^{N_f}m_{i}\int d^4 x
d^{2}\tht_{1} d^2\tht_{2}\frac{1}{\Psi^{2}_{3}}.
\eeq
When we consider the matter fields $Q$ and $\tilde Q$, as mentioned
above, this causes the change of prepotential:
\beq
{\cal F}=-\frac{1}{8\pi^2}\prod_{i=1}^{N_f}m_{i}
\frac{\La^{4-N_f}}{{\cal A}^2}.
\eeq 
This agrees with the result obtained in Ref.\cite{O}. \n
To summarize the result, we have derived one instanton
induced effective vertex
for the case of $N=2$ SQCD by the analogy of $N=1$ result.
The key idea used here is the behaviour of the $N=1$ propagator in the large
distance limit and its extention to the $N=2$ SUSY case.
We also used the $N=1,2$ SUSY promotion rule to derive the low energy 
effective lagrangian in the large distance limit.
This promotion rule is just within the classical level and
does not contain any quantum effect \cite{Y}. 
Under the proper normalization and the requirement of SUSY, the induced
vertex takes different form that depends on the VEV's of matter field.
It also depends on the representation of matter and here we concentrate
on the fundamental representation. \n
Finally let us briefly comment the VEV's of squarks.
From the previous analysis of $N=2$ SQCD vacuum structure, 
it is known that for non-zero quark masses, the VEV's of squarks are zero.
But when the quarks have  vanishing masses and 
for $N_{f}=0, 1$ case, the vacuum of the moduli develops only Coulomb branch
but for $N_{f}\ge 2$, it develops the so called Higgs branches where there
exist flat directions along which the gauge symmetry is completely broken
\cite{SW2,AL}. The explicit matrix form of these VEV's of squarks in moduli
space is discussed in \cite{AL}.
For further step in this direction along the work of Novikov \cite{N}
and Ito \cite{IS,IS3}, the generalization of this calculation to the group
$SU(N)$ or to an arbitrary Lie group has to be considered.

\vskip 1cm

\leftline{\bf Acknowledgement}

This work is supported in part by Ministry of Education (BSRI-96-2442), 
by KOSEF (961-0201-001-2) and by CTP/SNU through the SRC program of KOSEF.

\end{document}